\documentclass[aps,prl,superscriptaddress,showpacs]{revtex4}
\usepackage{graphicx}
\usepackage{enumerate}
\usepackage{eufrak}

\newcommand{\be}[1]{\begin{equation} \label{(#1)}}
\newcommand{\ee}{\end{equation}}
\newcommand{\ba}[1]{\begin{eqnarray} \label{(#1)}}
\newcommand{\ea}{\end{eqnarray}}

\newcommand{\barr}{\begin{array}}
\newcommand{\earr}{\end{array}}

\def\lsim{\raise0.3ex\hbox{$\;<$\kern-0.75em\raise-1.1ex\hbox{$\sim\;$}}}
\def\gsim{\raise0.3ex\hbox{$\;>$\kern-0.75em\raise-1.1ex\hbox{$\sim\;$}}}
\begin{document}

\title{Pinning down the mechanism of neutrinoless double beta
decay with measurements in different nuclei
}

\author{Frank Deppisch}
\email{frank.deppisch@desy.de}
\affiliation{Deutsches Elektronen-Synchrotron (DESY), D-22603
Hamburg, Germany}

\author{Heinrich P\"as}
\email{hpaes@bama.ua.edu}
\affiliation{Department of Physics \&
Astronomy, University of Alabama, Tuscaloosa, AL 35487, USA}

\begin{abstract}
A measurement of neutrinoless double beta decay in one isotope
does not allow to determine the underlying physics mechanism. We
discuss the discrimination of mechanisms for neutrinoless double
beta decay by comparing ratios of half life measurements for
different isotopes. Six prominent examples for specific new
physics contributions to neutrinoless double beta decay are
analyzed. We find that the change in corresponding ratios of half
lives varies from 60\% for supersymmetric models up to a factor of
5-20 for extra-dimensional and left-right-symmetric mechanisms.
\end{abstract}

\maketitle
%%%%%%%%%%%%%%%%%%%%%%%%%%%%%%%%%%%%%%%%%%%%%%%%%%%%%%%%%%%%%%

An uncontroversial detection of neutrinoless double beta
($0\nu\beta\beta$) decay  \cite{Doi:1982dn,revs,vogel,paesrev}
will be a discovery of uttermost significance. Most importantly,
it will prove lepton number to be broken in Nature, and neutrinos
to be Majorana particles \cite{petcov}. On the other hand, it will
immediately generate another puzzle: what is the mechanism that
triggers the decay? The most prominently discussed mechanism for
neutrinoless double beta decay is the exchange of light Majorana
neutrinos. But other mechanisms, like the exchange of SUSY
superpartners with R-parity violating or conserving couplings,
leptoquarks, right-handed W-bosons or Kaluza-Klein excitations,
among others, have been discussed in the literature as well.
Possibilities to disentangle at least some of the possible
mechanisms include the analysis of angular correlations between
the emitted electrons \cite{Doi:1982dn,Ali:2006iu} or a
comparative study of $0\nu\beta\beta$ and $0\nu\beta^+$ with
electron capture ($EC$) decay \cite{Hirsch:1994es}. Another
possibility seems to be the study of double beta decay to excited
$0^+$ states \cite{0+}. Unfortunately, the search for
$0\nu\beta^+/EC$ decay is complicated due to small rates and the
experimental challenge to observe the produced X-rays or Auger
electrons, and most double beta experiments of the next generation
are not sensitive to electron tracks or transitions to excited
states.

Without identification of the underlying mechanism, an
experimental evidence for neutrinoless double beta decay will only
provide ambiguous information about the concrete physics
underlying the decay. For example, no information about the
neutrino mass can be obtained from a measurement of the
neutrinoless double beta decay half life.

In general, contributions to neutrinoless double beta decay can be
categorized as either long-range or short-range interactions. In
the first case, the diagram involves two vertices which are
pointlike at the Fermi scale, and the exchange of a light neutrino
in between, and is described by an effective Lagrangian of the
type \cite{Pas:1999fc}
\begin{equation}
   {\cal L} = \frac{G_F}{\sqrt{2}}\left( j_{V-A}^{\mu}J_{V-A,\mu}+ \sum
              \epsilon_{NP} j_{NP} J_{NP}\right), \label{lrlag}
\end{equation}
where the  sum runs over all Lorentz invariant combinations of
hadronic and leptonic Lorentz currents of defined helicity,
$J_{NP,V-A}=\bar{u} {\cal O}_J d$ and $j_{NP,V-A} = \bar{e} {\cal
O}_j \nu$, respectively. Here ${\cal O}_{J,j} $ denotes the
corresponding transition operator. The effective coupling
strengths in new physics contributions are denoted as
$\epsilon_{NP}$ throughout. For short-ranged contributions, on the
other hand, the interactions are described by a single vertex
being pointlike at the Fermi scale. The decay rate therefore
results from first order perturbation theory, and is described by
the Lagrangian \cite{Pas:2000vn}
\begin{eqnarray}\label{lagsr}
   {\cal L} &=& \frac{G^2_F}{2} m_p^{-1} \sum \epsilon_{NP} J_{NP}
                J_{NP} j'_{NP}.
\end{eqnarray}
Here $m_p$ denotes the proton mass and the sum runs over all
Lorentz invariant combinations of hadronic,
$J_{NP}=\overline{u}{\cal O}_J d $, and leptonic,
$j'_{NP}=\overline{e}{\cal O}_{j} e^C$, currents of defined
chirality.

The combination involving two vertices of the first term in
(\ref{lrlag}) leads to the usual neutrinoless double beta decay
half life formula for the mass mechanism,
\begin{equation}
   [T_{1/2}^{m_\nu}]^{-1}=
   (\langle m_\nu \rangle/m_e)^2 G_{01}|{\cal M}^{m_\nu}|^2,
\end{equation}
where $\langle m_\nu \rangle$ is the effective neutrino mass in
which the contributions of individual neutrino mass eigenstates
are weighted by mixing matrix elements squared, $\langle m_\nu
\rangle = |\sum U_{ei}^2 m_i|$. The combination of the first term
in (\ref{lrlag}) with any of the latter terms as well as the
short-range Lagrangian (\ref{lagsr}) leads to the expression
\begin{equation}\label{t12np}
   [T_{1/2}^{NP}]^{-1}=\epsilon_{NP}^2 G_{NP} |{\cal M} ^{NP}|^2.
\end{equation}
Here, ${\cal M}^{m_\nu}$ and ${\cal M}^{NP}$ are the nuclear
matrix elements for the mass mechanism and alternative new physics
contributions, and $G_{01}$ and $G_{NP}$ denote the corresponding
phase space integrals from the list given in \cite{Doi:1982dn}. We
have assumed, that one mechanism dominates the double beta decay
rate, and we do not consider interference between different
mechanisms. Calculational details and results for the relevant
matrix elements involved have been given elsewhere
\cite{Pas:1999fc,Pas:2000vn}, and numerical results for all common
double beta emitter isotopes will be published soon \cite{paes06}.

In the present context, we will concentrate on the observation
that the combinations of leptonic and hadronic currents specific
to different mechanisms result in different nuclear matrix
elements. This fact taken alone is not of much help in order to
disentangle the  different mechanisms, since e.g. a smaller
nuclear matrix element for the mass mechanism as compared to any
alternative new physics mechanism can be compensated by a larger
value for the neutrino mass, at least within the constraints
implied by other observations such as Tritium beta decay and
cosmology. However, under the assumption that one mechanism
dominates in triggering the decay, the new physics parameter
$\langle m_\nu \rangle$ or $\epsilon_{NP}$ drops out in the ratio
of experimentally determined half lives for two different emitter
isotopes,
\begin{equation}
   \frac{ T_{1/2}(^AX)}{T_{1/2}(^{76}{\rm Ge})} =
   \frac{|{\cal M}(^{76}{\rm Ge})|^2 G(^{76}{\rm Ge})}
   {|{\cal M}(^AX)|^2 G(^AX)}.
\end{equation}
Consequently, half life ratios depend on the mechanism of double
beta decay, but not on the new physics parameter, and thus can be
compared with the theoretical prediction for different mechanisms.
Moreover, the error in the isotope nuclear matrix element ratio
can be reduced compared to the theoretical error in one matrix
element, due to cancellations of systematic effects.

In the following we study several prominent examples of specific
alternative new physics contributions by calculating the
corresponding ratios of half lives
\begin{equation}
   {\cal R}^{NP}(^AX)=\frac {T_{1/2}^{NP}(^AX)}
   {T_{1/2}^{NP}(^{76}{\rm {\rm Ge}})}, \label{ratiosdef}
\end{equation}
where we concentrate on a comparison with $^{76}$Ge as it
constitutes the best tested isotope to date. We choose the
following mechanisms for a detailed discussion:

\begin{itemize}

\item{\bf SUSY-accompanied neutrinoless double beta decay:
${\cal R}^{\rm SUSYacc}$}\\
This mechanism has been first discussed in \cite{susyacc}. The
effective Lagrangian for the dominant contribution assumes the
form
\begin{eqnarray}
   {\cal L}&\supset&
   \frac{G_F U^*_{e i}}{4 \sqrt{2}}\epsilon^{\rm SUSYacc}
   \Big[ \left( \overline{\nu}_i (1+\gamma_5) e^c\right)
   \left( \overline{u} (1+\gamma_5) d \right)
%   \nonumber \\ &&
   + \frac{1}{2} \left( \overline{\nu}_i \sigma^{\mu\nu}
   (1+\gamma_5) e^c\right)
   \left( \overline{u} \sigma^{\mu \nu}(1+\gamma_5) d \right)
   \Big],
\end{eqnarray}
and results from integrating out a heavy $d$-squark of the $k$-th
generation with $R$-parity violating couplings $\lambda'_{11k}$
and $\lambda'_{1k1}$, and exchanging a light neutrino of the
$i$-th generation between the nucleons. The new physics parameter
is given by
\begin{equation}
   \epsilon^{\rm SUSYacc} =
   \sum_k\frac{\lambda'_{11k}\lambda'_{1k1}}{2 \sqrt{2} G_F} \sin 2
   \theta_k \left(\frac{1}{m^2_{\tilde{d}_1}}- \frac{1}
   {m^2_{\tilde{d}_2}} \right),
\end{equation}
where $\theta_k$ parametrizes the left-right sfermion mixing of
the mass eigenstates $\tilde{d}_1$ and $\tilde{d}_2$.

\item{\bf Gluino exchange mechanism in R-parity
violating SUSY}: ${\cal R}^{{\rm SUSY}-\tilde{g}}$\\
In this short-range contribution discussed in
\cite{Mohapatra,Hirsch:1995ek}, integrating out $u$- and
$d$-squarks and a gluino leads to the effective Lagrangian
\begin{eqnarray}
   {\cal L}&\supset&
   \frac{G_F^2}{2} m_p^{-1} \epsilon^{\tilde{g}}
   \left( (\overline{u} (1+\gamma_5) d) (\overline{u} (1+\gamma_5) d)
   - \frac{1}{4} (\overline{u} \sigma^{\mu \nu} (1+\gamma_5) d)
   (\overline{u} \sigma^{\mu \nu} (1+\gamma_5) d) \right)
%   \nonumber \\ &&
(\overline{e}(1+\gamma_5) e^c),
\end{eqnarray}
with
\begin{equation}
   \epsilon^{\tilde{g}}=
   \frac{2 \pi \alpha_s}{9}
   \frac{\lambda'^2_{111}}{G_F^2 m^4_{\tilde{d}_R}}
   \frac{m_p}{m_{\tilde{g}}}
   \left[1+\left(\frac{m_{\tilde{d}_R}}{m_{\tilde{u}_L}}\right)^4
\right].
\end{equation}

\item{\bf Right-handed currents}: ${\cal R}^{LR-\eta\eta}$ and
${\cal R}^{LR-\lambda\lambda}$\\
Integrating out right-handed $W$-bosons occurring in left-right
symmetric models can lead to two types of new contributions with
right-handed leptonic currents \cite{Doi:1982dn},
\begin{equation}
   {\cal L} \supset \frac{G_F}{\sqrt{2}}
   \left(\overline{\nu}_i\gamma_\mu(1+\gamma_5) e^c\right)
   \Big(\eta (\overline{u}\gamma^\mu (1-\gamma_5) d)
   + \lambda (\overline{u} \gamma^\mu(1+\gamma_5) d) \Big),
\end{equation}
where the new physics parameters are given by $\eta$ and $\lambda$.

\item{\bf Kaluza-Klein neutrino exchange in extra-dimensional
models}: ${\cal R}^{KK}$\\
In extra-dimensional theories, the double beta observable is given
by a sum over contributions from all Kaluza-Klein excitations with
masses $m_{(n)}$, weighted with the mass dependent matrix element
${\cal M}^{m_{\nu}}(m_{(n)})$ \cite{Bhattacharyya:2002vf}:
\begin{equation}
   \epsilon^{KK} =
   \frac{1}{{\cal M}^{m_{\nu}}}\sum_{-\infty}^{\infty} U^2_{en} m_{(n)}
   \left({\cal M}^{m_{\nu}}(m_{(n)}) -{\cal M}^{m_{\nu}}\right).
\end{equation}
In this case the effective coupling constant $\epsilon^{KK}$
depends on the nuclear matrix element ${\cal
M}^{m_{\nu}}(m_{(n)})$, and therefore the particle physics does
not decouple from the nuclear physics. This is because the masses
of the Kaluza-Klein excitations vary from values much smaller than
the nuclear Fermi momentum $p_F$ to values much larger than $p_F$,
while the $m_{(n)}$-dependence of ${\cal M}^{m_{\nu}}(m_{(n)})$
changes around $p_F$. Therefore the Kaluza-Klein spectrum has to
be fixed by choosing specific values for the brane shift parameter
$a$ and the radius of the extra dimension $R$. In the limit of
$a\rightarrow 0$ or $R\rightarrow 0$, ${\cal R}^{KK}$ approaches
${\cal R}^{m_\nu}$.

\end{itemize}

The matrix elements for the mass mechanism and for the SUSY
accompanied neutrino exchange have been calculated in the pn-QRPA
approach of \cite{Staudt:1990qi,Hirsch:1994es}, in the latter case
for the first time. For the other mechanisms, existing numerical
values obtained with the same nuclear structure model have been
adopted from the literature. The values for the phase space
integral factors $G_{01}$, $G_{NP}$ have been calculated in
\cite{Doi:1982dn}. Numerical values for ${\cal R}^{NP}(^AX)$ are
given in Table 1, and Fig.~1 displays the relative change expected
from various new physics contributions, compared  to the mass
mechanism. An application of the procedure to any other
alternative new physics contribution by using the matrix elements
listed in \cite{paes06} is straightforward.

All isotope ratios have been normalized to the half life of the
most extensively studied nucleus $^{76}$Ge. Moreover, while at
present no experiment using a $^{128}$Te source has been proposed,
we included this isotope since it provides a particularly powerful
discriminator and thus may encourage future experimental efforts
to study this nucleus.

\begin{table}[!t]
\begin{center}
\begin{tabular}{l|cccccc|c}
\hline
& $^{82}$Se & $^{100}$Mo & $^{128}$Te & $^{130}$Te & $^{136}$Xe & $^{150}$Nd & Ref. \\
\hline
${\cal R}^{m_\nu}$ & 0.26 &  0.11 &  3.26 &  0.18 &  0.77 &
0.02 & this paper
\\
${\cal R}^{\rm SUSYacc}$ & 0.28 &  0.11 &  3.22 &  0.17 &  0.53  &
0.02  & this paper
\\
${\cal R}^{{\rm SUSY}-\tilde{g}}$ & 0.28 &  0.10 &  3.16 &  0.17 &
0.53 &   0.01 & \cite{Hirsch:1995ek}
\\
${\cal R}^{LR-\eta\eta}$ & 0.29 &  0.13 &  2.96 &  0.20 &  0.54 &
0.02 & \cite{Muto:1989cd}
\\
${\cal R}^{LR-\lambda\lambda}$ & 0.14 &  0.13 &  18.40 &   0.13 &
0.67 &  0.01 & \cite{Muto:1989cd}
\\
${\cal R}^{KK}$ (10 GeV$^{-1}$) & 0.24 &  0.08 &   3.26 &   0.19 &
3.31 &  0.08 & \cite{Bhattacharyya:2002vf}
\\
${\cal R}^{KK}$ (0.1 GeV$^{-1}$) & 0.26 &  0.11 &   3.26 &   0.18
&   0.78 &   0.02 & \cite{Bhattacharyya:2002vf}
\\
\hline
\end{tabular}
\caption{ Ratios ${\cal R}(^AX)$ of half lives for various
important double beta decay emitter isotopes, normalized to the
half-life of $^{76}$Ge. For the exchange of Kaluza-Klein
excitations in extra dimensional theories the brane shift
parameter and bulk radius do not factorize, and are chosen to be
$a=10~ {\rm GeV}^{-1},~~ 0.1~ {\rm GeV}^{-1}$ and
$R=(1/300)$~eV$^{-1}$.}
\end{center}
\end{table}

\begin{figure}[!t]
\centering
\includegraphics[clip,width=0.95\textwidth]{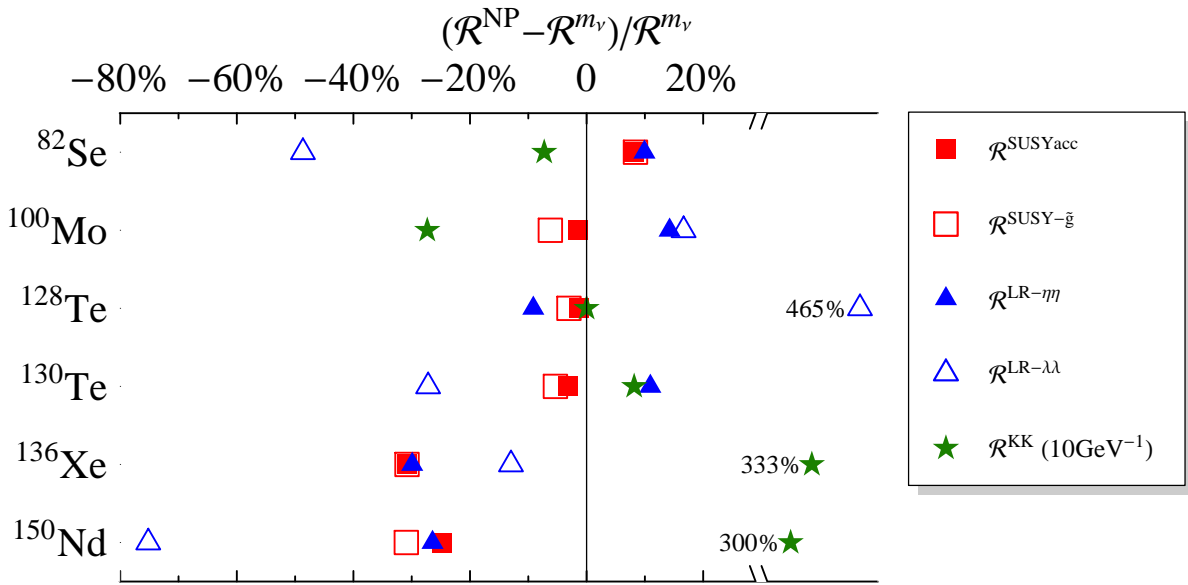}
\caption{Relative deviations of half life ratios ${\cal
R}^{NP}(^AX)$, normalized to the half-life of $^{76}$Ge,
compared
to the ratio in the mass mechanism ${\cal R}^{m_\nu}(^AX)$. }
\label{bulkpath}
\end{figure}

The two supersymmetric contributions show similar deviations,
which are rather small for all isotopes. It is obvious that these
mechanisms are most effectively discriminated from the mass
mechanism by comparing the half life ratios between $^{82}$Se and
$^{136}$Xe which vary by 60\%. In left-right symmetric models,
strong deviations can be found for the $\lambda\lambda$
combinations, while deviations for the $\eta\eta$ combination are
rather small. A comparison of half life ratios between $^{100}$Mo
and $^{136}$Xe yields a variation of 70 \% for the $\eta\eta$
contribution with right-handed hadronic currents, while a
comparison of measurements in $^{128}$Te and $^{150}$Nd will
provide a powerful discriminator with a variation of more than a
factor of 20 for the $\lambda\lambda$ contribution with
left-handed hadronic currents. Similarly in extra-dimensional
neutrino models with a large brane shift parameter, large
deviations can be found for $^{136}$Xe and $^{150}$Nd, and the
half life ratios for $^{150}$Nd and $^{100}$Mo vary by more than a
factor of 5. Some caution is necessary when referring to the half
life ratio of the heavily deformed $^{150}$Nd, which is ignored in
most QRPA calculations (compare the discussion in \cite{def}).
Finally it should be stressed that not necessarily two positive
results are needed - already the comparison of one half life
measurement and one upper bound in another isotope could provide
non-trivial information on the double beta mechanism.

Since the theoretical errors of the nuclear matrix element
calculation dominate the experimental errors, it is difficult to
determine the confidence level with which either mechanism can be
excluded to generate the observed double beta evidence. If, for
example, a statistical distribution of matrix element values is
assumed, a relative variation of 60\% in ${\cal R}^{NP}(^AX)$ with
respect to ${\cal R}^{m_\nu}(^AX)$ is significant only if the
corresponding nuclear matrix elements would be known with an
accuracy of 15\%, which seems to be unrealistic, if only one pair
of isotopes is being analyzed. Indeed, estimates of errors in
nuclear matrix elements vary from a factor 3-5, when the spread of
published values is used as a measure, to only 30\%, according to
an assessment of uncertainties inherent in QRPA \cite{unc}.

However, the significance of the comparison of two isotopes will
increase if a whole set of measurements in different isotopes
resembles the expected pattern. Moreover, one would expect that
systematical effects, like an overestimation of the nuclear matrix
elements due to a too small value for the particle-particle
interaction $g_{pp}$ in the pn-QRPA approach, a different value
for the axial-vector coupling $g_A$, the inclusion of higher-order
terms or a different model-space would influence calculations for
the different isotopes in a similar way, and thereby cancel in the
half life ratios discussed. This expectation is confirmed by the
comparison of the results of different QRPA codes in \cite{unc},
and of QRPA and shell model codes in \cite{vogel}. Finally it has
been pointed out in \cite{bilenky} that the half life ratios
(\ref{ratiosdef}) can also be used to single out the correct
nuclear structure model. In this case the correct combination of
mechanism and nuclear structure code can be determined by the best
fit of the theoretical half life ratios to half life measurements
in various nuclei. Thus the results presented in this letter
should be complemented and checked with alternative codes for the
nuclear matrix element calculation. Moreover, other mechanisms,
including pion exchange \cite{pion}, may be dominating in some of
the models discussed, and should be discussed as well.

In summary, we discussed how different mechanisms of neutrinoless
double beta decay would manifest themselves in half life ratios
involving different isotopes. We thus conclude that complementary
measurements in different isotopes would be strongly encouraged.
At present, next-generation experiment proposals exist for
$^{76}$Ge (GERDA, MAJORANA, GEM, GeH$_4$), $^{82}$Se (Super-NEMO,
DCBA, SeF$_6$), $^{100}$Mo (MOON), $^{130}$Te (CUORE), $^{136}$Xe
(EXO, XMASS, Xe), as well as for the isotopes $^{48}$Ca,
$^{116}$Cd and $^{160}$Gd not discussed in this letter (CANDLES,
COBRA and GSO) (for recent overviews of the experimental status 
see \cite{exps}). An experimental study of this kind should be
complemented by neutrino mass searches in Tritium beta decay
experiments and cosmology, as well as searches for effects of the
alternative new physics source of lepton number violation in other
processes, such as lepton flavor violating decays
\cite{Cirigliano:2004tc}.

After this paper had been submitted for publication, the paper
\cite{Gehman:2007qg} appeared, which comes to similar conclusions
and estimates the number of required measurements and their
precision needed.

\section*{Acknowledgements}
HP thanks B. Allanach, K.S. Babu and S. Pascoli for
discussions and
the University of Hawaii and DESY for kind
hospitality.

\end{document}